\documentclass{iau}
\usepackage{graphicx,natbib}
 
\newcommand{\lambdaRe}{$\lambda_{\rm Re}$}

\title{Angular Momentum across the Hubble sequence from the CALIFA survey}

\author[Falc\'on-Barroso, Lyubenova \& van de Ven]{Jes\'us Falc\'on-Barroso$^{1,2}$,
Mariya Lyubenova$^{3,4}$, Glenn van de Ven$^4$\\ and the CALIFA collaboration}

\affiliation{$^1$Instituto de Astrof\'isica de Canarias, V\'ia L\'actea s/n, La Laguna, Tenerife, Spain\\
$^2$Departamento de Astrof\'isica, Univ. de La Laguna, E-38205 La Laguna, Tenerife, Spain. email: {\tt jfalcon@iac.es}\\
$^3$Kapteyn Astronomical Institute. University of Groningen Postbus 800, 9700 AV Groningen, The Netherlands. email: {\tt lyubenova@astro.rug.nl}\\
$^4$Max Planck Institute for Astronomy, K\"onigstuhl 17, 69117 Heidelberg, Germany. email: {\tt glenn@mpia.de}}

\pubyear{2014}
\volume{311}
\jname{Galaxy Masses as Constraints of Formation Models}
\editors{M. Cappellari \& S. Courteau, eds.}

\begin{document}

\maketitle

\begin{abstract}
We investigate the stellar angular momentum of galaxies across the Hubble
sequence from the CALIFA survey. The distribution of
CALIFA elliptical and lenticular galaxies in the \lambdaRe\,$-$\,$\epsilon_{\rm e}$ diagram is
consistent with that shown by the Atlas$^\mathrm{3D}$ survey. Our data, however, show that
the location of spiral galaxies in this diagram is significantly different. We
have found two families of spiral galaxies with particularly peculiar
properties: (a) spiral galaxies with much higher \lambdaRe\ values than any elliptical and lenticular galaxy;
(b) low-mass spiral galaxies with observed \lambdaRe\ values much lower than
expected for their apparent flattening. We use these two families of objects to
argue that (1) fading alone cannot explain the transformation of spiral to lenticular galaxies, 
and (2) that those low-mass spiral galaxies are in fact dark matter dominated, which explains the unusually low angular momentum.
\end{abstract}

\firstsection

\section{The CALIFA survey}

The CALIFA survey (S\'anchez et al.\ 2012) provides a morphologically unbiased
and statistically well defined view of the stellar and ionised gas properties of
up to 600 galaxies in the redshift range $0.005<z<0.03$.  The long wavelength
coverage (3400--7300\,\AA) allows an accurate determination of the stellar
population and ionised gas properties (Marino et al.\ 2013, P\'erez et al.\
2013).  

In the work we present here we use the dedicated set of high-spectral resolution
observations over 3750--4550\,\AA\ with the PPAK V1200 grating to derive
high-quality stellar kinematics for a sample of 300 galaxies from elliptical E
to spiral Sd morphological types. The wide field-of-view of the PMAS/PPAK
instrument routinely covers well beyond the effective radius of the observed
galaxies.

\section{Angular momentum across the Hubble sequence}

\begin{figure}
\centering
\includegraphics[width=0.6\columnwidth]{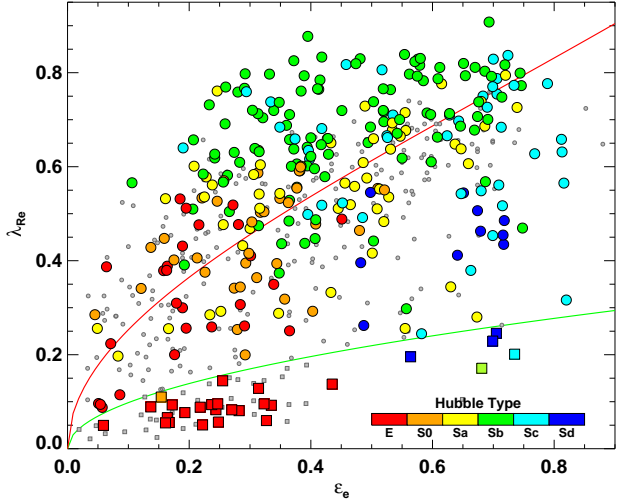}
\caption{Apparent stellar angular momentum \lambdaRe\ as function of the
ellipticity $\epsilon_{\rm e}$ for 300 CALIFA galaxies color-coded by their
morphological type. The green curve indicates the demarcation line between
slow-rotator (squares) and fast- rotator galaxies (circles) as inferred from the
Atlas$^\mathrm{3D}$ survey of elliptical (E) and lenticular (S0) galaxies, which
are plotted as grey symbols (Emsellem et al.\ 2011). The red curve indicates the
location in case galaxies were simple isotropic oblate rotators viewed edge-on.}
\label{fig:fig1}
\end{figure}

The combination of the apparent specific stellar angular momentum \lambdaRe\ and
the ellipticity $\epsilon_{\rm e}$ (Fig.~\ref{fig:fig1}) is an often used
diagnostic tool to constrain the dynamical structure and evolution of galaxies.
If all galaxies would be simple oblate isotropic rotators they should follow the
red curve or lie to the left of it when viewed away from edge-on. The green
curve indicates the demarcation line between fast rotator (FR, circles) and slow
rotator (SR, squares) galaxies as inferred from the Atlas$^\mathrm{3D}$ survey
of elliptical (E) and lenticular (S0) galaxies (grey symbols, Emsellem et al.
2011).\medskip

With the CALIFA survey we are exploring for the first time in a homogeneous way
all Hubble types. Spiral galaxies, as expected, are nearly all fast rotator
galaxies. However, in comparison with E and S0 galaxies, especially the Sb
galaxies reach very high \lambdaRe\ values, while several Sc and Sd galaxies
exhibit low \lambdaRe\ values and fall just inside the slow rotator regime, in
both cases having interesting consequences as discussed in the following
sections.

\section{Are lenticular galaxies faded spirals?}

The angular momentum is set very early on in the life of galaxies, providing an
important tool to distinguish between different scenarios of galaxy evolution.
For example, a possible explanation of the observed morphology-density relation
is the transformation of spiral into fast rotating lenticular galaxies through
fading of the stellar populations (e.g. Cappellari et al.\ 2011). Under this
scenario the light concentration and the angular momentum are not expected to
change significantly.\medskip

\begin{figure}
\begin{center}
\includegraphics[width=0.75\columnwidth]{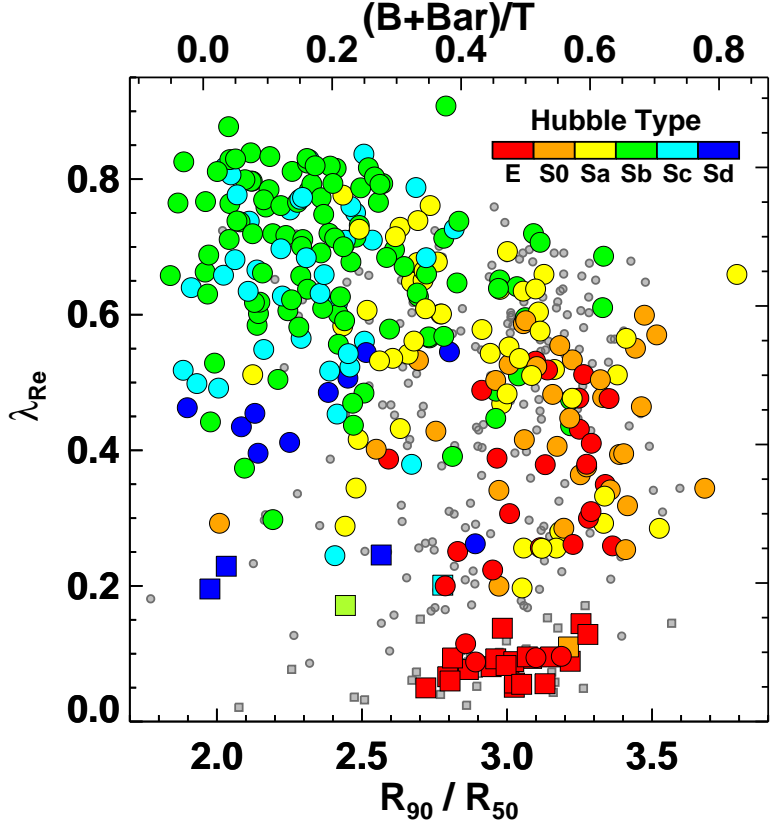}
\end{center}
\caption{Apparent stellar angular momentum \lambdaRe\ versus light concentration
$R_{90}/R_{50}$ based on SDSS $r$-band Petrosian radii (bottom axis) as a proxy
for the relative flux in a spheroid (bulge+bar) and disk component (top axis).
Symbols  and colors as in Fig.\ref{fig:fig1}.}
\label{fig:fig2}
\end{figure}

We measured the apparent specific stellar angular momentum within the half-light
radius \lambdaRe\ (left axis in Fig.~\ref{fig:fig2}) for our sample of 300
CALIFA galaxies. The light concentration based on SDSS $r$-band Petrosian radii
(bottom axis) is a proxy for the relative flux in a spheroid (bulge+bar) and
disk component (indicated in the top axis, following Gadotti 2009). We observe
that Sa galaxies indeed overlap with S0/FR galaxies.  Sc and Sd galaxies also
have low \lambdaRe\ values, but their concentrations are too small, which taken
together with their large gas supplies, makes it difficult to transform them
into red sequence S0/FR galaxies by fading alone.  Most of the Sb galaxies are
in the green valley and hence their gas may have already been quenched. However,
their \lambdaRe\ values are twice as large and light concentrations
significantly smaller than S0/FR galaxies today.  This hints to a picture in
which the red sequence is being built over time from galaxies that contain disks
that are substantially larger and have higher angular momenta.

\section{Are low-mass spiral galaxies dark matter dominated?}

\begin{figure}
\centering
\includegraphics[width=\textwidth]{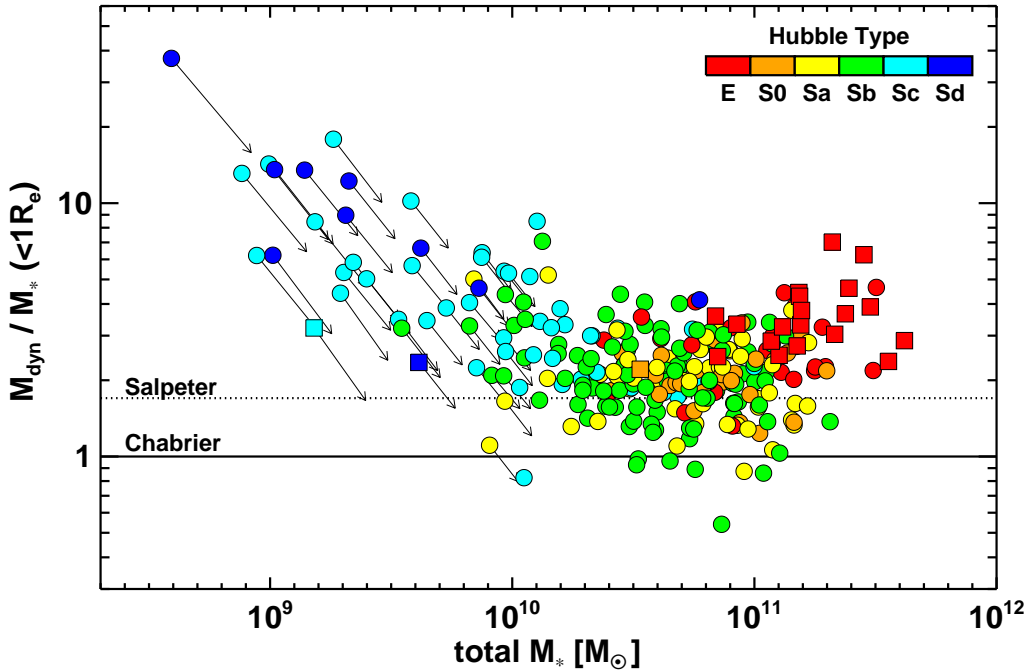}
\caption{Ratio of total dynamical mass over stellar mass within the half-light
radius versus total stellar mass for our sample of CALIFA galaxies. Horizontal
lines mark normalizations for different stellar initial mass functions. Symbols
and colors as in Fig.\ref{fig:fig1}.}
\label{fig:fig3}
\end{figure}

The intrinsic mass distribution of a galaxy is one of the most consequential
properties when it comes to its structure and evolution. However, the interplay
between baryons and dark matter content is still under heated debate. We
inferred the total mass of the 300 CALIFA galaxies of all morphological types by
constructing axisymmetric dynamical Jeans models that fit the observed motions
of their stars. The stellar mass was derived by fitting the galaxy spectra to
stellar population models under the assumption of a Chabrier initial stellar
mass function (IMF; Gonz\'alez-Delgado et al. 2014). The resulting ratio of
total-to-stellar mass inside the effective radius (R$_{\rm e}$) is shown in
Fig.~\ref{fig:fig3} (vertical axis), with arrows indicating the approximate
contribution of gas to the baryonic mass (Papastergis et al. 2012). On the
high-mass end, the upturn above unity could be due to the presence of dark
matter or a change toward an IMF even bottom-heavier than the Salpeter IMF as
claimed in previous studies (e.g., Cappellari et al. 2012, La Barrera et al.
2013). However, for low-mass Sc-Sd galaxies there are no indications of a
significant deviation from a Chabrier IMF, so that the high total-to-stellar
mass ratios are believed to be the result of a high dark matter fraction even
within R$_{\rm e}$. This is further supported by their observed low \lambdaRe\
values, as the presence of a relatively large dark matter halo would support a
dynamically hot but geometrically thin stellar disk.

\section*{Acknowledgements}

\noindent
The authors are grateful to the organizers for the very stimulating meeting
celebrating such a special occasion. We are also indebted to the rest of the
CALIFA collaboration for the strong support and assistance to carry out this
work. JFB acknowledges support from grants AYA2010-21322-C03-02 and
AIB-2010-DE-00227 from the Spanish Ministry of Economy and Competitiveness
(MINECO). JFB and GvdV acknowledge support from FP7 Marie Curie Actions of the
European Commission, via the Initial Training Network DAGAL under REA grant
agreement number 289313.

\end{document}